\documentclass[amsmath,amssymb,twocolumn]{article}

\usepackage{graphicx}

\usepackage{bm}
\usepackage{graphicx}
\usepackage{amsmath,amssymb,amsfonts}
\usepackage{subfigure}
\usepackage{verbatim}

\newcommand{\vect}[1]{\boldsymbol{#1}}
\newcommand{\bvect}[1]{\bar{\boldsymbol{#1}}}

\newcommand{\bsigma}{\boldsymbol{\sigma}}

\graphicspath{{images/}}

\begin{document}

\twocolumn[{

\centerline{\LARGE On the net displacement of contact surface centroid in contractile bodies}
\vspace{3ex}

\centerline{Jos\'e J. Mu\~noz${}^{1,2}$, Lucie Condamin${}^3$, David Doste${}^4$}

\vspace{3ex}



 {${}^1$Mathematics Department, Laboratori de C\`alcul Num\`eric (LaC\`aN), Universitat Polit\`ecnica de Catalunya, Barcelona, Spain.}
 
    {\tt http://www.lacan.upc.edu/jose.munoz, }{ \tt j.munoz@upc.edu}
    
 {${}^2$Centre International de M\`etodes Num\`erics en Enginyeria (CIMNE), Barcelona, Spain.}
 
{%
${}^3$Institut National des Sciences Appliqu\'ees, Lyon, France.
}%

{${}^4$Facultat de Matem\`atiques i Estad\'istica, Universitat Polit\`ecnica de Catalunya, Barcelona, Spain.}


\begin{abstract}
We investigate the motion of the contact surface centroid for contractile bodies on substrates with a viscous friction law and when inertial forces are negligible. We deduce a set of sufficient conditions that ensure that the surface centroid remains still. The conditions are automatically satisfied for linear analysis and homogeneous constant viscous friction parameters. In non-linear analysis additional requirements are necessary: i) the material is incompressible, ii) the material points in contact do not vary, and iii) the surface is flat.  These results demonstrate the inability of slender organisms to move under homogeneous viscous contact condition if the contact surface remains constant, regardless of the contractility strategy employed. We numerically simulate some situations that do not comply to these conditions, such as the use of non-homogeneous or anisotropic friction, which illustrate possible strategies for net propulsion.
\end{abstract}

\textbf{Keywords:}
centre of mass, locomotion, worm, friction, viscosity

\vspace{3ex}
}]





\section{Introduction}

Mechanisms for locomotion of organisms have been well studied in fluids \cite{cohen10,stone96} and granular media \cite{avron04,texier17}. However, only in the former case exist general principles for propulsion, which have been mainly derived for low Reynolds number \cite{lauga09,shapere89}. When bodies are submerged in a fluid, Purcell's scallop theorem furnishes sufficient conditions for the null net motion in fluids through the analysis of the parametric space of body configurations \cite{purcell77,shapere89}. In these results, Stokes equations are imposed, with a no-slip boundary condition. 

Limbless self-propulsion of deformable solids on loose soil or granular matter has been analysed and modelled in detail \cite{astley20,goldman14}, and a theory that allows determining the propulsion from the interaction with the granular environment, so-called resistive force theory, has been derived and validated \cite{gray55,maladen09,goldman14}. Moreover, optimal strategies for locomotion and cost reduction have been successfully deduced and simulated \cite{noselli14,jiang17,tanaka12}. However, partially due to the complex interaction at the boundary, no general equivalent theorem furnishing sufficient conditions for the no net motion of the body centroid (centre of mass when density is constant) on a frictional substrate has been established. This paper aims at furnishing some results along this direction.

\begin{figure}[!htb]
  \centering
  \includegraphics[width=0.3\textwidth]{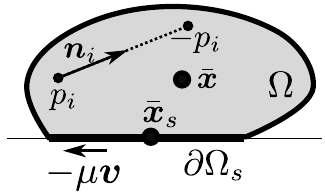}
  \caption{Schematic of a self contractile body $\Omega$ subjected to viscous friction at the bottom boundary and with contractile dipole $\pm p_i\vect n_i$. Large dots indicate body centroid $\bar{\vect x}$ and contact surface centroid ${\vect x}_s$.}
  \label{f:body}
\end{figure}

In order to analyse the locomotion on a viscous frictional substrate, we assume a body $\Omega$ with the ability to self-contract and in contact with the substrate in a region $\partial\Omega_s$, as shown in Figure \ref{f:body}. Similar to the analysis in fluids, and due to the small size of $\Omega$ and the low accelerations, we neglect the contribution of inertial effects \cite{hu09}. Eventually, we will consider thin or elongated organisms, such that in this case boundary forces can be approximated by body forces applied in the body domain $\Omega$, which is initially occupying domain $\Omega_0$. We will also comment the situation when the whole limbless body is permanently in contact with the frictional substrate. The lifting or removal of friction will be discussed after deriving our main results.

\section{Motion of contact surface centroid}

We are interested in analysing the motion of the body centroid $\bvect x=\int_{\Omega} \vect x dV/V$, with volume $V=\int_{\Omega}  dV$, and the motion of the contact surface centroid $\bvect x_s=\int_{\Omega} \vect x dS/S$, with $S=\int_{\partial\Omega_s}  dS$. We will also denote by $\vect x$ and $\vect x_0$ the position of a material point of body $\Omega$ in its current deformed position and in its reference configuration, respectively.

The ability to  \emph{self-contract} is represented by a field of $N$ dipoles with opposing forces $\vect f=\sum_i^N\pm p_i\vect n_i\delta(\vect x^\pm_i)$ and magnitude $p_i$, applied at points $\vect x=\vect x_i^+$ and $\vect x=\vect x_i^-$ along directions $\pm\vect n_i=\pm(\vect x_i^+-\vect x_i^-)/||\vect x_i^+-\vect x_i^-||$, respectively. Note that the set of dipoles is indeed self-equilibrated, i.e.
\begin{align}\label{e:selfe}
\int_\Omega \vect f dV=\sum_i ^N \pm p_i \vect n_i=\mathbf 0.
\end{align}

 The static equilibrium of domain $\Omega$ is then given by Cauchy's equation, including the dipoles as a body force \cite{bonet97}:
\begin{align}\label{e:cauchy}
\nabla\cdot\bsigma  + \vect f = \mathbf 0,\ \forall \vect x \in \Omega,
\end{align}
and with boundary conditions that we assume stemming from a viscous frictional condition
\begin{align}\label{e:BC}
\begin{aligned}
\bsigma\vect n&=-\mu \vect v, &&\forall \vect x\in \partial \Omega_s, \\
\bsigma\vect n&=\vect 0, &&\forall \vect x\in \partial \Omega\backslash \partial \Omega_s,
\end{aligned}
\end{align}
%
%
with $\vect v=\dot{\vect x}$ the material velocity. The associated weak form (or virtual principle) to problem \eqref{e:cauchy}-\eqref{e:BC} follows after pre-multiplying equation \eqref{e:cauchy} by a set of compatible virtual displacements $\vect \delta\vect u$, integrating by parts and considering the boundary condition in \eqref{e:BC}:
\begin{align*}
&\text{\emph{Find $\vect u$ such that, for all virtual admissible}}\\
&\text{\emph{ displacements $\vect\delta \vect u$}},
\end{align*}
\begin{align}\label{e:wf}
&\int_\Omega\!\! \vect\varepsilon(\vect \delta\vect u):\vect\sigma(\vect u) dV =&&
\pm\sum_i  \vect\delta\vect u\cdot \vect f_i^{\pm} (\vect x_i^\pm)\nonumber \\
&&& - \int_{\partial \Omega_s} \!\!\!\mu\vect\delta\vect u \cdot \vect v dS,
\end{align}
with $\vect\varepsilon(\vect \delta\vect u)$ the virtual strains. In linear analysis, $\vect\varepsilon(\vect \delta\vect u)=\frac{1}{2}\left(\nabla \vect \delta\vect u + (\nabla \vect \delta\vect u)^T\right)$, while in finite strains and large displacements, $\vect\varepsilon(\vect \delta\vect u)=\mathbf F^{-T}\vect \delta \mathbf E\mathbf F^T$, with $\mathbf E=\frac{1}{2}\left(\mathbf F^T\mathbf F-\mathbf I\right))$  the Lagrangian or Green strain tensor, and the deformation gradient $\mathbf F$ defined by $\mathbf F=\frac{\partial\vect x}{\partial \vect x_0}$ \cite{bonet97}. 

The equality in \eqref{e:wf} must hold for all compatible virtual displacements $\vect\delta\vect u$, which in our case includes a constant displacement $\vect \delta\hat{\vect u}$. Therefore, since $\vect\varepsilon(\vect \delta\hat{\vect u})=\vect 0$ and the dipoles are self-equilibrated (see equation \eqref{e:selfe}), equation \eqref{e:wf} implies the following relation,
\begin{align}\label{e:v0}
\int_{\partial \Omega_s} \mu\vect v dS=\vect 0.
\end{align}

In linear analysis, where $dS\approx dS_0$, and with constant homogeneous viscosity,  this equation implies that the position of the contact surface centroid $\bar{\vect x}_s$ is constant. In order to show that in general non-linear analysis this is not so, we recall Nanson's formula \cite{holzapfel00}, $dS\vect n=J\ dS_0\mathbf F^{-T}\vect N$, with $(dS, \vect n)$ and $(dS_0, \vect N)$ the surface differential and normal vectors in the deformed and reference configuration, respectively, and $J=det(\mathbf F)$.  By using also the relations $\dot J=(\nabla\cdot\vect v) J$ and $\dot{\mathbf F}^{-1}=-\mathbf F^{-1}\nabla\vect v$, it then follows that the time variation of the surface differential can be expressed as,
\begin{align}\label{e:dds}
\dot{dS} &=(\nabla\cdot\vect v)dS+JdS_0\left(\dot{\vect n}-\nabla\vect v\vect n\right)\cdot \mathbf F^{-T}\vect N\nonumber \\
&=(\nabla\cdot\vect v)dS
+dS\left(\dot{\vect n}-\nabla\vect v\vect n\right)\cdot\vect n.
\end{align}

As a consequence,
\begin{align}\label{e:dxs}
\dot{\bar{\vect x}}_s=&
\int_{\partial \Omega_{s}} \vect v dS 
+\int_{\partial \Omega_{s}} \vect x (\nabla\cdot \vect v)dS  \nonumber \\
&+ \int_{\partial \Omega_{s}} \vect x\left(\dot{\vect n}-\nabla\vect v\vect n\right)\cdot\vect ndS.
\end{align}

From this expression, we can deduce a set of conditions that together, ensure the motion of the contact surface centroid vanishes:
\begin{itemize}
\item[\emph{i)}] The friction coefficient $\mu$ is constant and homogeneous.
\item[\emph{ii)}] The material is incompressible.
\item[\emph{iii)}] The material points that are in contact with the substrate remain in contact.
\item[\emph{iv)}] The substrate is flat.
\end{itemize}

Conditions \emph{i)} and \emph{ii)} are very common in many \emph{in vitro} situations in biomechanics. These, together with the result in \eqref{e:v0}, imply that the first two terms in the rhs of \eqref{e:dxs} vanish. As a consequence of condition \emph{iii)},  $\vect n\cdot \partial_n\vect v=\vect n\cdot\nabla \vect v\vect n= \vect 0$, that is, the material region $\partial\Omega_s$ is constant, while condition \emph{iv)} yields  $\dot{\vect n}=\vect 0$. Summarising, when conditions \emph{i)-iv)} are satisfied, we have that, 
\begin{align}\label{e:xs0}
 \dot{\bar{\vect x}}_s=\vect 0.
\end{align}

\section{Discussion}

In deriving the result in \eqref{e:xs0} we have determined sufficient kinematic and frictional conditions for the null motion of the contact surface centroid. We recognise that some of them were expected, such as the requirement of homogeneous and constant friction on flat surfaces, or having no variations of the normal velocity across the contact surface, so that the contact surface is maintained. However, other particular conditions that may give rise to contact surface net motion, such as compressibility or using a non-flat surface, even when viscosity is constant and homogeneous, are less evident. 

Considerable work has been devoted to the analysis of sufficient conditions for locomotion of immersed bodies, where organisms are subjected to friction or no-slip conditions on the whole surface \cite{lauga09,purcell77}. However, a similar analysis of the sufficient conditions for no net motion when the body is partially in contact with a viscous substrates has not been analysed, to the authors' knowledge, although some of the sufficient conditions for locomotion have been well studied \cite{astley20,cohen10}. 

For instance, for flat or elongated shapes, and when nearly the whole body is in contact with the substrate, body and contact surface centroids are approximately  equivalent, i.e. $\bar{\vect x}_s\approx \bar{\vect x}$. Net motion of $\bar{\vect x}$ may be then achieved by varying the contact surface in different regions of the boundary, like in sidewinding \cite{astley20}. Our first example in the next section simulates such situation.

We emphasise that relation \eqref{e:v0} shows that the surface centroid does not move in linear analysis and when $\mu=const>0$. For more general contact friction laws with the form $\vect\sigma\vect n=-\vect\mu(\vect v)$, the result in  equation \eqref{e:v0} must be replaced by,
\begin{align}\label{e:uv0}
\int_{\partial\Omega} \vect\mu(\vect v) dS=\vect 0.
\end{align}

This relation does not imply the immobility of $\bar{\vect x}_s$, and in consequence, diversions from the homogeneous constant frictional condition may give rise to changes in $\bar{\vect x}_s$. Some examples are the use of different parallel and normal frictional coefficients, as employed in the resistive force theory  \cite{sharpe15,gray55,maladen09}, or resorting to different coefficients in different tangential directions, as it is the case of euglenoids in fluids \cite{arroyo12}.

We point out that relation \eqref{e:xs0} is valid for any  contractile strategy and material constitutive law.  For the particular case of elastic bodies, the result in \eqref{e:xs0} can be interpreted as the solution of minimising a total energy functional. For showing this, we first discretise the weak form in \eqref{e:wf} with a time-stepping $\vect v_{n+1}\approx (\vect u_{n+1}-\vect u_n)/(t_{n+1}-t_n)$. The displacement at time $t_{n+1}$, denoted by $\vect u_{n+1}$ is then equivalent to finding the minimiser of a functional $W(\vect u_{n+1})$ given by
\begin{align*}
W(\vect u_{n+1})&=\int_\Omega \phi(\vect u_{n+1}) dV
+\sum_i  p_i l_i(\vect u_{n+1}) \\
& + \frac{1}{2(t_{n+1}-t_n)} \int_{\partial\Omega} \mu ||\vect u_{n+1}-\vect u_n ||^2 dS,
\end{align*}
with $l_i(\vect u)=||\vect x_i^+-\vect x_i^-||$ the length of the dipole, and $\phi(\vect u)$ an elastic energy density function. By noting that both $l_i(\vect u)$ and $\phi(\vect u)$ are invariant with respect to rigid body motions, and splitting the displacement $\vect u_{n+1}$ into the surface mean value $\bar{\vect u}_s$ and the deviation $ \Delta \vect u_s$, that is, by  writing $\vect u_{n+1}= \bar{\vect u}_s+ \Delta \vect u_s$, with $\bar{\vect u}_s=\int_{\partial\Omega} \vect u_{n+1} dS/S$, it can be shown that with the same conditions \emph{i)-iv)}, we have that 
\begin{align}\label{e:ineqs}
W(\vect u_{n+1})=W(\Delta \vect u_s) +\mu ||\bar{\vect u}_s ||^2 S.
\end{align}

Since the solution $\vect u_{n+1}$ minimises $W(\vect u_{n+1})$, the surface mean value must vanish, i.e. $\bar{ \vect u}_s=\vect 0$.

\section{Numerical Examples}

\begin{figure}[htb!]
  \centering
  \includegraphics[width=0.24\textwidth]{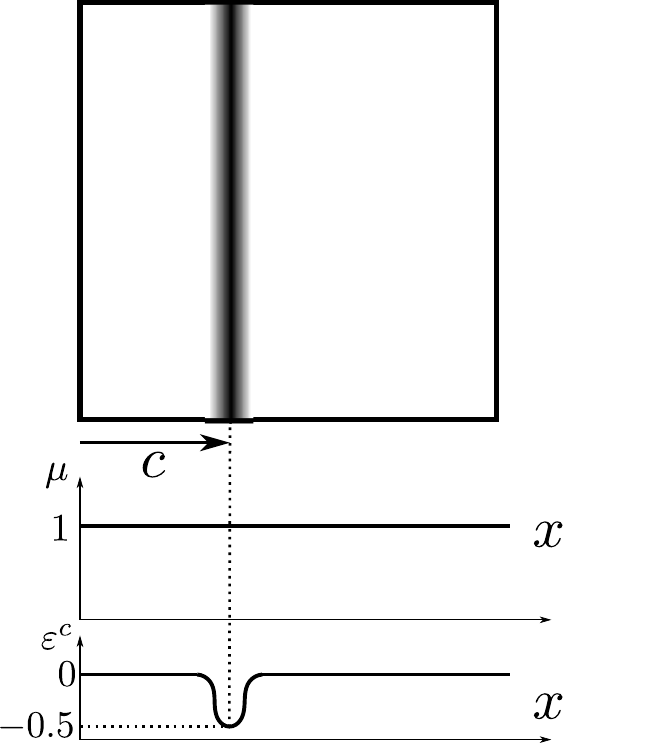}\hspace{-3ex}
  \includegraphics[width=0.24\textwidth]{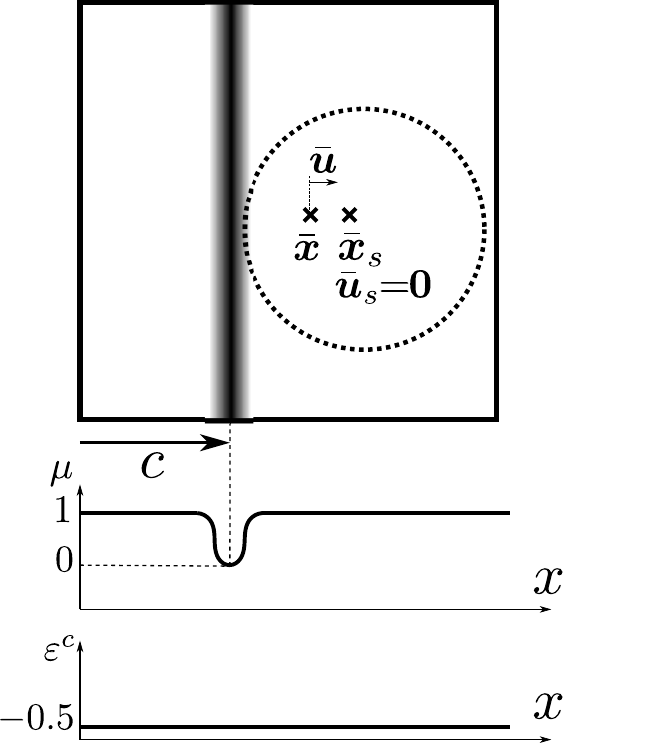}\\
    Case a)\hspace{20ex}Case b)
  \caption{Example 1: Contractile square domain. a) Homogeneous friction with a localised contractility on a vertical band. b) Homogeneous contractility with a localised reduction of friction on a vertical band. }
  \label{f:Square1}
\end{figure}

We illustrate our results with a couple of examples that will be solved numerically. The first one consists on a flat square domain $[0,L]^2$ in contact with a substrate, and contracting according to the following two strategies: a) constant friction everywhere with an applied local contractility along a thin vertical domain  (see Figure \ref{f:Square1}a), and b) a contraction on the whole domain with a null friction along a thin vertical line (see Figure \ref{f:Square1}b). We test different horizontal positions $c$ of the contracting band in case a) and different positions of the frictionless band in case b). The problem is solved resorting to a finite element discretisation and the contraction is implemented by superimposing to the elastic deformation an isotropic contractile strain $\varepsilon^c\mathbf I$.

\begin{figure}[htb!]
  \centering
  \includegraphics[width=0.18\textwidth]{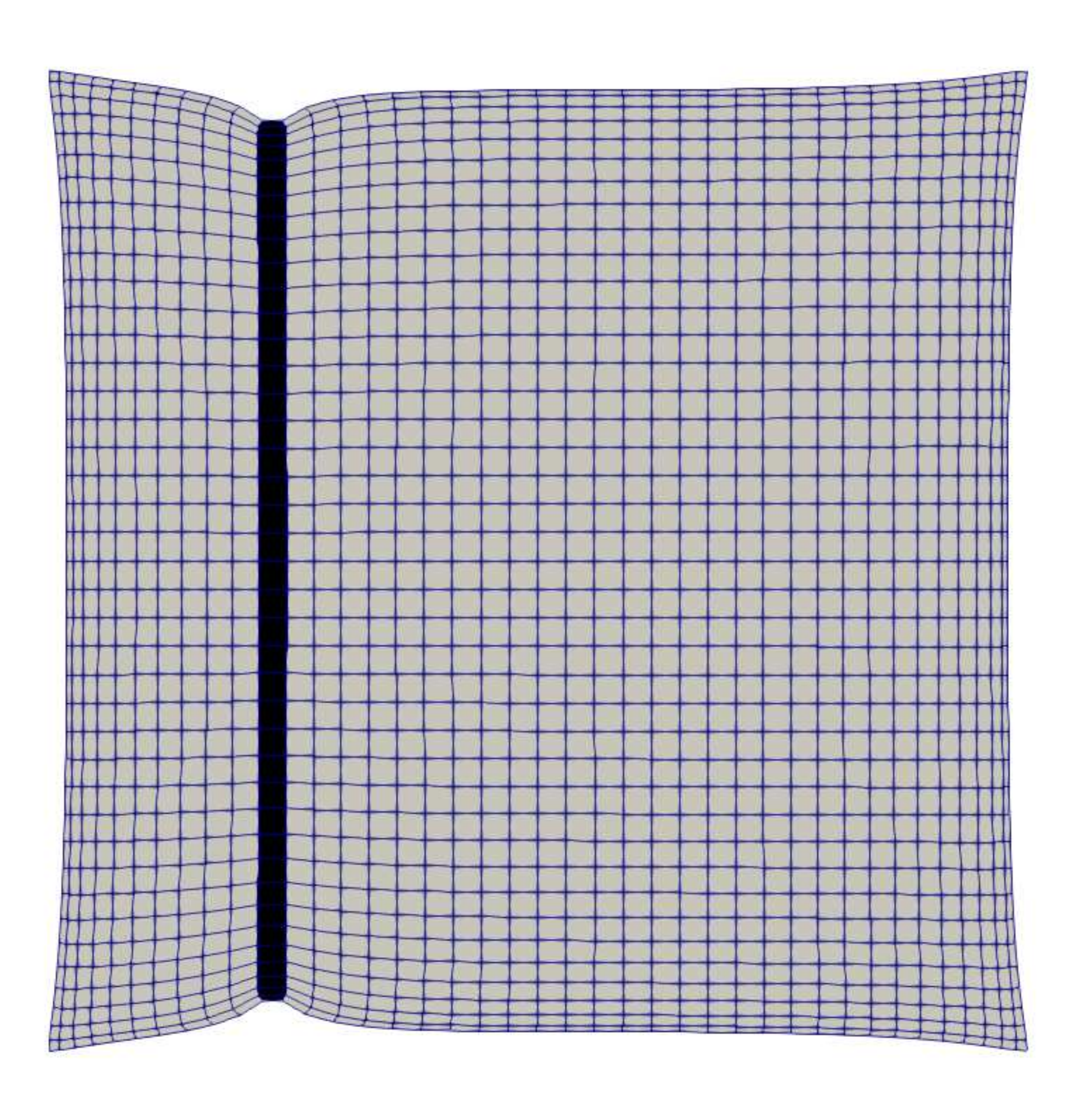}
  \includegraphics[width=0.28\textwidth]{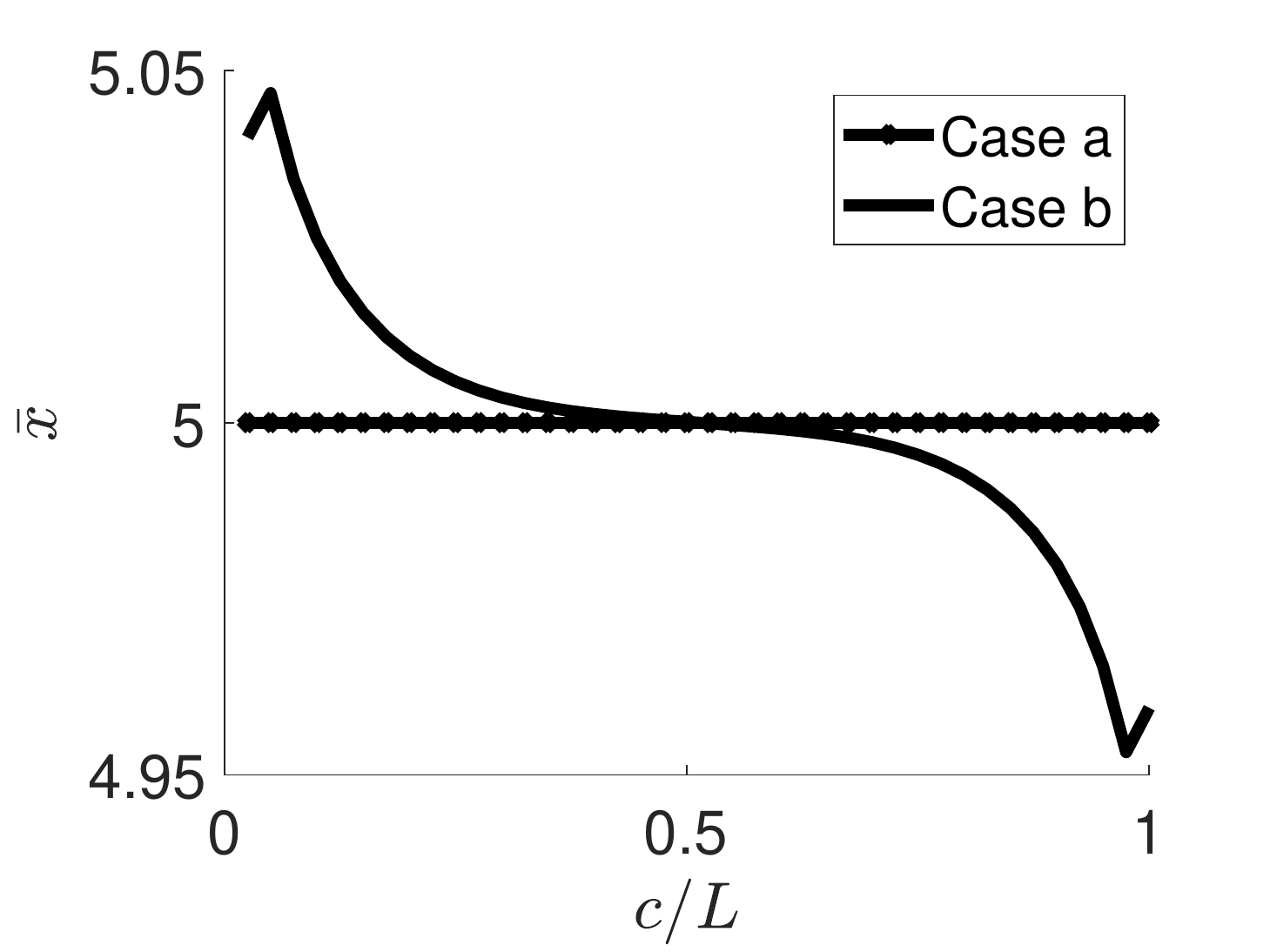}
  \caption{Example 1: Left: deformation for case b. Right: resulting displacement of motion of horizontal coordinate of the body (surface) centre. }
  \label{f:Square2}
\end{figure}

 The numerical results confirm that in both cases the displacement of the centroid of the surface that has non-zero friction vanishes, i.e.  $\bar{\vect u}_s=\vect 0$, for any contractile strategy in case a), and for each one of the friction distributions in case b). We show in Figure \ref{f:Square2} the plot of $\bar{\vect u}$ for different locations of the contractile band in case a) and for each location of the frictionless band in case b). In case a), since the frictional surface also coincides with the whole body surface, we have that $\bar{\vect u}=\vect 0$ for all values of $c$. Instead, in case b), since the centre of the frictional surface is moving backwards as $c$ increases, the centre of the whole body $\bar{\vect  x}$ moves forwards for $c<L/2$, while $\bar{\vect x}$ moves backwards for $c>L/2$. When $c=L/2$, the two centres coincide, $\bar{\vect x}_s=\bar{\vect x}$, and in this case no net displacement of the body centre is obtained, i.e. $\bar{\vect u}=\bar{\vect u}_s=\vect 0$. 

\begin{figure}[htb!]
  \centering
  \includegraphics[width=0.18\textwidth]{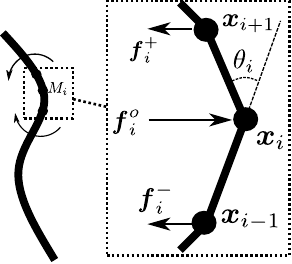}
  \includegraphics[width=0.28\textwidth]{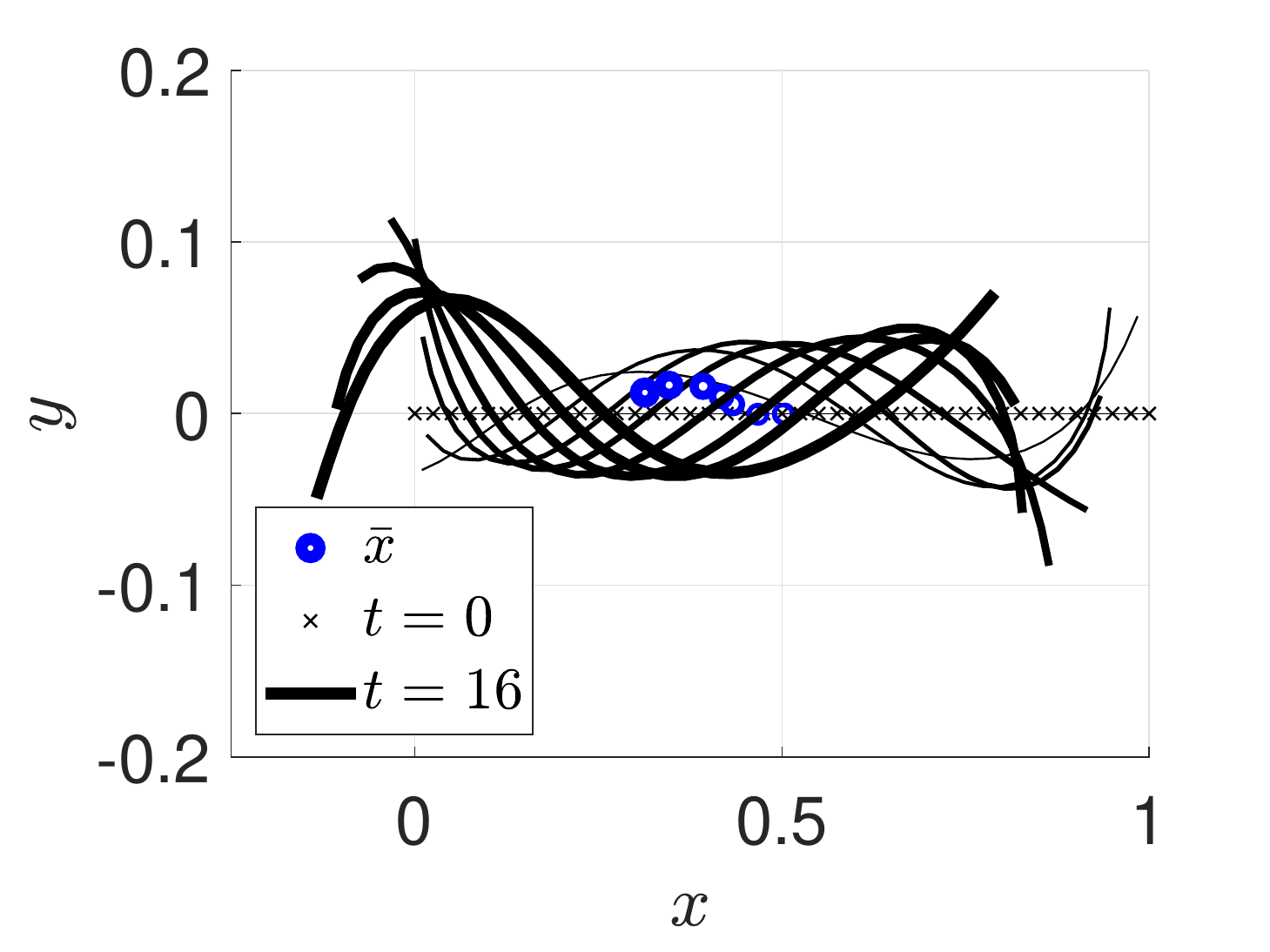}
  \caption{Example 2: Left: Worm-like domain with self-contractile moments. Right: Snapshots of deformed shape for anisotropic friction ($\mu_n=10\mu_t=1$).}
  \label{f:worm1}
\end{figure}

In a second example, we analyse a worm-like unidimensional domain that has the ability to bend and produce a wave-like undulation, similarly to the analysis presented for hydrodynamics in \cite{taylor51}. The worm is modelled here as a set of $n$ planar line segments joining $n+1$ nodes $\vect x_0,\ldots , \vect x_n$. Stretching and bending elasticity is furnishing by using the following total elastic potential
\begin{align*}
\Phi(\vect x)=\sum_{i=0}^n k(||\vect x_{i+1}-\vect x_i||-l_{i0})^2 + \sum_{i=1}^n k_\theta \sin^2\theta_i,
\end{align*}
with $l_{i0}$ the initial length of segment $\vect x_{i+1}-\vect x_i$ and $\theta_i$ the angle between $\vect x_{i+1}-\vect x_i$ and $\vect x_{i}-\vect x_{i-1}$ (see Figure \ref{f:worm1}). In addition to the elastic forces, the model is subjected to nodal frictional components  $\vect f_i^\mu=-\mu \vect v_i$ and a set of bending moments $M_i$ applied at the interior nodes $i=1,\ldots, n-1$. Each bending moment $M_i$ is in fact decomposed into three self-equilibrated forces $\vect f_i^-, \vect f_i^o, \vect f_i^+$, respectively applied at nodes $\vect x_{i-1}, \vect x_i$ and $\vect x_{i+1}$. These forces are normal to direction $\vect x_{i+1}-\vect x_{i-1}$, and such that $\vect f_i^- + \vect f_i^o + \vect f_i^+=\mathbf 0$, with a resulting bending moment equal to $M_i$, as also illustrated in Figure \ref{f:worm1}(left). More specifically, the relation between $M_i$ and the three nodal forces is given by:
\begin{small}
\begin{align}\label{e:Mi}
0&=f_i^- + f_i^o + f_i^+\\
0&=\vect e_z\cdot (f_i^-(\vect x_{i-1}-\vect x_i)
+f_i^-(\vect x_{i+1}-\vect x_i))\times \vect n_i\nonumber\\
M_i&=\vect e_z\cdot (f_i^-(\vect x_{i-1}-\vect x_i)
-f_i^-(\vect x_{i+1}-\vect x_i)\times\vect n_i)\times \vect n_i
\nonumber
\end{align}
\end{small}
with $\vect e_z=\{0,\ 0,\ 1\}^T$ and $\vect n_i=\vect e_z\times(\vect x_{i+1}-\vect x_{i-1})/||\vect x_{i+1}-\vect x_{i-1}||$ the approximated normal vector at node $i$. These conditions define uniquely the direction and magnitude of the forces for a given value of $M_i$. The equilibrium equations for each node $i$ read then,
\begin{align}\label{e:wbe}
\frac{\partial \Phi(\vect x)}{\partial\vect x_i} + \vect f^M_i=\vect f_i^\mu,
\end{align}
where vector $\vect f^M_i=\vect f_{i-1}^+ +\vect f_i^o + \vect f_{i+1}^-$ includes all the force contributions at node $i$ due to the contractile bending moments. We have applied a distribution of time varying moments equal to
\[
M_i=\sin(\omega t - k s_i),
\]
except at the end points, where $M_0=M_n=0$. Here, $\omega$ is the frequency, $k$ the wave number, and $s_i$ the initial position of point $\vect x_i$. The discretised weak form of the balance equations in \eqref{e:wbe} reads,
\begin{align*}
&\text{\emph{Find $\vect u_i$ such that for all virtual
 displacements $\vect\delta\vect u_i$}}\\
&\vect\delta\vect u_i\cdot\frac{\partial \Phi(\vect x)}{\partial\vect x_i} + \vect\delta\vect u_i\cdot\vect f^M_i=\vect\delta\vect u_i\cdot\vect f_i^\mu.
\end{align*}
 
Since this relation must be also satisfied for arbitrary rigid body displacements, and we also have that $\sum_i \vect f^M_i=\vect 0$ (forces due to applied moments are self-equilibrated by construction, according to \eqref{e:Mi}) and $\sum_i\frac{\partial\Phi}{\partial \vect x_i}=\vect 0$, we have an equivalent result to the one in \eqref{e:v0}:
\begin{align*}
\sum_i \mu\vect v_i=\vect 0.
\end{align*}

For a constant viscous coefficient, this result implies that the mean velocity also vanishes. We have numerically solved the set of $n$ non-linear equations in \eqref{e:wbe}, and verified that when isotropic friction is used, the resulting motion of the worm centre $\dot{\bar{\vect x}}=\sum_i \vect v_i /n$ is exactly zero, up to machine tolerance. When the same oscillatory set of moments is used, but in conjunction with anisotropic friction $\vect f^\mu=-\mu_t \vect v_t - \mu_n\vect v_n$, with $\vect v_t$ and $\vect v_n$ the tangential and normal components of the displacements, and $\mu_t$ and $\mu_n$ two different friction coefficients, the worm is propelled, as shown in Figure \ref{f:worm1} (right). In this case, the worm centre moves in the opposing direction of 	the wave of bending moments, as also pointed out for locomotion in fluids  \cite{cohen10,lauga09} and solids \cite{goldman14}.

\begin{figure}[htb!]
  \centering
(a)  \includegraphics[width=0.4\textwidth]{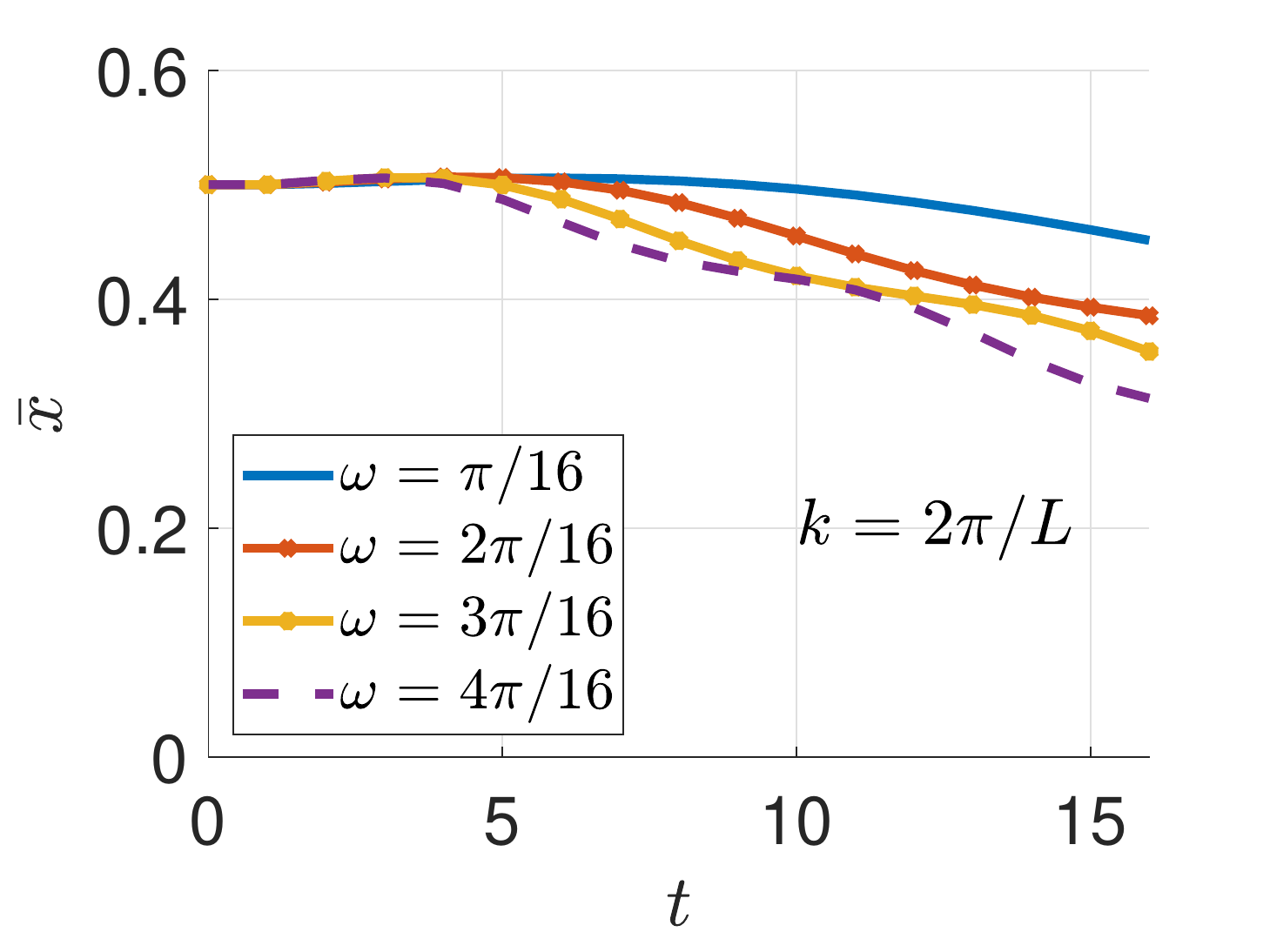} \\
(b)  \includegraphics[width=0.4\textwidth]{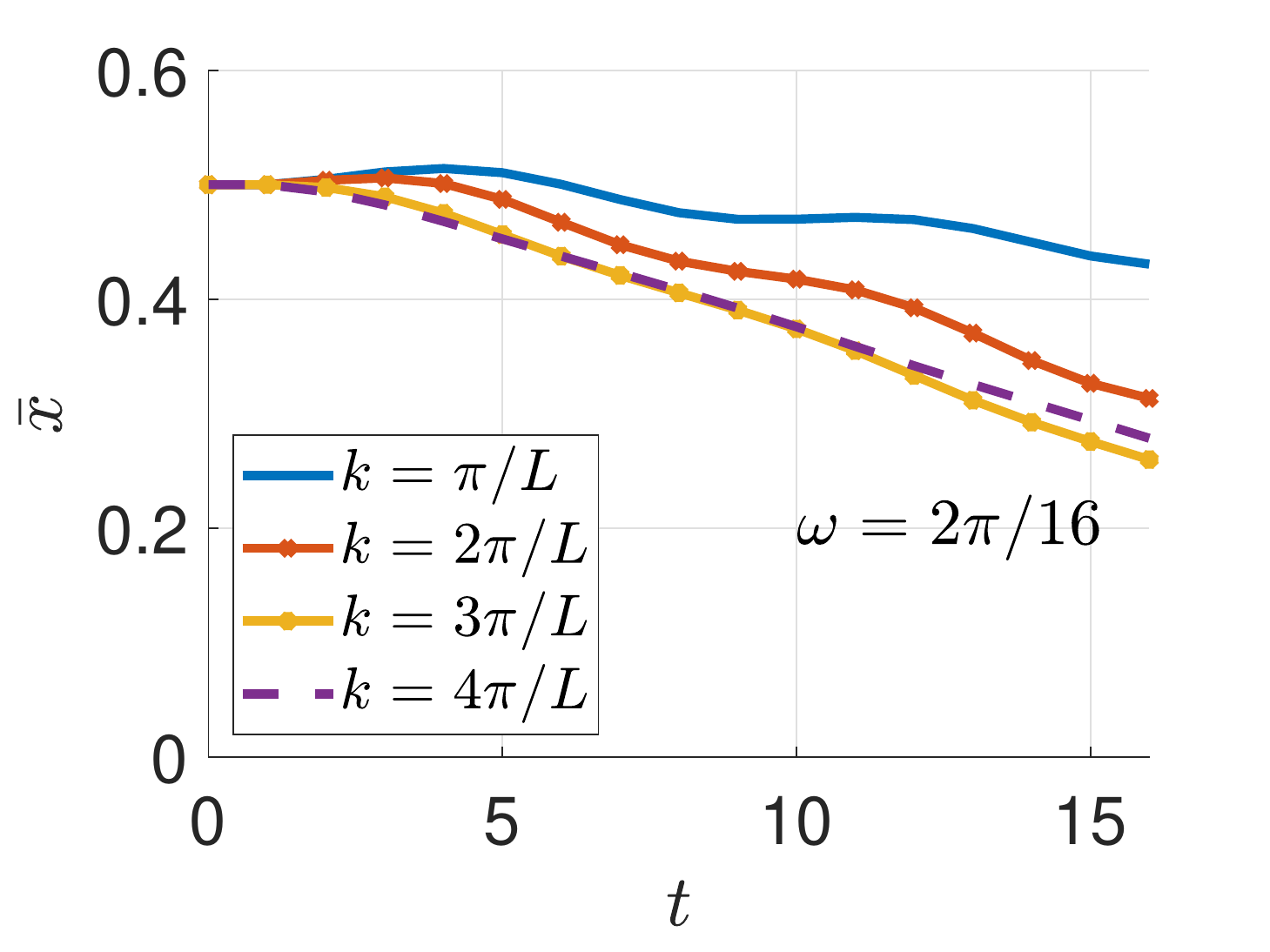}
  \caption{Example 2. (a) Horizontal displacement of the worm centroid for increasing values of frequency. (b) Horizontal displacement of the worm centroid for increasing values  and wave number . }
  \label{f:worm2}
\end{figure}

We have tested different frequencies $\omega$ and wave numbers $k$. The sequence of horizontal displacements shown in Figure \ref{f:worm2} indicates that although final displacement increases for increasing values of $\omega$ and $k$, there is for both parameters a limit value beyond which no substantial improvement is observed. This fact is also confirmed by the low wave numbers and limited frequencies that organisms generally exhibit \cite{goldman14}. We have not focused this study on the search of optimal modes or strategies, also with respect to other motion parameters such as amplitude or friction, which have been analysed elsewhere \cite{avron04,kano14,vanleeuwen15,hatton13}, or simulated with the discrete element method \cite{ding12}. We intend to analyse in future works more realistic resistive forces such as frictional plasticity models  \cite{askari16}, or more complex waves such as those in sidewinding \cite{astley15} or helical motion \cite{texier17}, and interpret them as a drift from the homogeneous frictional conditions where no motion is achieved.

\section{Conclusions}

We have derived a set of sufficient conditions for maintaining the contact surface centroid at rest. The two examples presented above exploit the violations of some of these conditions, giving rise to net propulsion.  The first example uses a reduction of friction in some regions of the contact surface, while the second example resorts to anisotropic friction. These are in contrast with some well-known  conditions required for achieving locomotion in fluids, such as the non-time reversibility of the body deformations \cite{cohen10,purcell77,lauga09}.

In one of our numerical examples we obtain propulsion by reducing friction on a subdomain of the body surface. The resulting centroid displacement is similar to the one studied in fluids when a contractile wave. In our case though, the net movement is not the result of alternating vortices \cite{lauga09}, but rather a difference between the positions of the body centroid $\bar{\vect x}$, and the contact surface centroid $\bar{\vect x}_s$.  

We remark that our theoretical results are applicable to solids with arbitrary constitutive laws and contractile strategies, and applicable to large displacements and deformations. However, our conclusions cannot be generalised to bodies immersed in a fluid, where a no-slip condition is assumed in this case. 

We also mention that our analysis is pertinent to crawling microorganisms on frictional substrates such as \emph{C. elegans} on agar \cite{stephens08,cohen10}. In many instances, net locomotion is achieved through the activation of contractile waves, as it is the case in lateral undulation and sidewinding, with some lifting/lowering, which results in a variation of the frictional conditions. While the measurement of the centre of mass in swimming organisms \cite{xiong14} and human gait dynamics \cite{tesio10} is common, its analysis and correlation with substrate friction at cell mechanics is usually not investigated.

Indeed, understanding sufficient conditions for the null net movement may help to elucidate the conditions for migration in cells or monolayers on substrates, where planar motion is dominant \emph{in vitro}. For instance, substantial computational and experimental efforts  have been devoted  to traction force microscopy (retrieval of traction field exerted by cells or tissues) \cite{butler02,alamo13,legant13,sunyer16}. In those cases, the motion of the tissue centre should be interpreted as a set of unequal adhesions, and not solely to differential tissue contractility. Accordingly, the configurational parametric space should be analysed jointly with modes of the contact conditions.


\bibliographystyle{plain}


\end{document}